\documentclass[10pt, conference]{IEEEtran}

\ifCLASSINFOpdf
\else
\fi
\usepackage{amsmath,amssymb,bm}
\usepackage{mathtools,nccmath}
\usepackage{algorithmic}

\usepackage{array}
\usepackage{tikz}
\usepackage{tikzscale}
\usepackage{pgf}
\usepackage[T1]{fontenc}
\usepackage{mathtools}
\usepackage{bbm}
\usepackage[numbers,sort&compress]{natbib}
\usepackage[linesnumbered,lined,ruled,commentsnumbered]{algorithm2e}
\usepackage{lipsum}  
\usepackage{authblk}
\usepackage{graphicx}
\usepackage{subcaption}

\IEEEoverridecommandlockouts

\begin{document}
%
\title{Performance of Slotted ALOHA in User-Centric Cell-Free Massive MIMO }
\vspace{-0.25cm}
%

\author[1,2]{Dick~Maryopi}
\author[1]{Daud~Al~Adumy}
\author[3]{Osman~Musa}
\author[4,5]{Peter~Jung}
\author[1]{Agus~Virgono}

\affil[1]{\normalsize School of Electrical Engineering, Telkom University, Bandung, Indonesia.}
\affil[2]{\normalsize The University Center of Excellence for Advanced Intelligent Communications (AICOMS), Telkom University, Indonesia.}
\affil[3]{\normalsize BIFOLD, Technical University of Berlin, 10587 Berlin, Germany}
\affil[4]{\normalsize Communications and Information Theory Group, Technical University of Berlin, 10587  Berlin, Germany}
\affil[5]{\normalsize German Aerospace Center (DLR) 
\vspace{-0.25cm}
\thanks{This work was supported in part by Telkom University No. 447/ PNLT3/ PPM/ 2022 and in part by  the Technical University of Berlin.}}

 \makeatletter
 \def\ps@IEEEtitlepagestyle{%
   \def\@oddfoot{\mycopyrightnotice}%
   \def\@evenfoot{}%
 }
 \def\mycopyrightnotice{%
   {\footnotesize Accepted for presentation at IEEE PIMRC 2024. Copyright may be transferred without notice, after which this version may no longer be accessible.\hfill}
   \gdef\mycopyrightnotice{}
 }

%




\maketitle

\begin{abstract}
To efficiently utilize the scarce wireless resource, the random access scheme has been attaining renewed interest primarily in supporting the sporadic traffic of a large number of devices encountered in the Internet of Things (IoT). In this paper we investigate the performance of slotted ALOHA\textemdash a simple and practical random access scheme\textemdash in connection with the grant-free random access protocol applied for user-centric cell-free massive MIMO. More specifically, we provide the expression of the sum-throughput under the assumptions of the capture capability owned by the centralized detector in the uplink. Further, a comparative study of user-centric cell-free massive MIMO with other types of networks is provided, which allows us to identify its potential and possible limitation. Our numerical simulations show that the user-centric cell-free massive MIMO has a good trade-off between performance and fronthaul load, especially at low activation probability regime.


\end{abstract}

\begin{IEEEkeywords}
Cell-free, Massive MIMO, Random Access, ALOHA, massive machine-type communication, Internet of Things.
\end{IEEEkeywords}

%
\IEEEpeerreviewmaketitle

\section{Introduction}
%

Over the past decade, there has been a pressing need to connect a massive number of devices to the wireless infrastructure to support a wide range of applications of IoT including massive machine-type communication (mMTC) and ultra reliable low latency communication (URLLC). While 5G wireless networks have envisioned connecting 1 million devices per km$^2$, the number will be pushed up to 10 million in the upcoming 6G networks \cite{9040264}. Beyond the large number, the main characteristics of this type of communication are the sporadic traffic, short block length, and low-power consumption, which have made its system design rather challenging \cite{9060999, 9205230}. Because only a small fraction of a large number of users are active at irregular intervals, dedicating a particular resource to all users would be an inefficient solution. In this case, the conventional multiple-access scheme might not be suitable.

A popular approach to address the problem is the so-called \textit{grant-free} random access protocol \cite{6843162,8340053, 7996932}. Instead of waiting for permission and following a complicated long procedure, in the grant-free protocol, the active users can immediately send their data packets on a given shared resource. The simple yet common scheme used for random access is the slotted ALOHA \cite{1093713}. When any user has a packet to transmit, he can directly transmit the packet in a given slot, regardless of other users transmitting their packets simultaneously. Any possible collision can be treated as packet loss and then resolved with packet retransmission according to a certain policy. In this fashion, the latency can be expected to reduce and the resource can be more efficiently utilized. 

Further, it is increasingly becoming appealing to provide wide wireless coverage and high data rate to enable critical and broadband IoT applications. However, as most IoT devices have limited power, this can become another serious problem along with a large number of devices. Primarily in the uplink direction, the transmit power of IoT devices is typically very low which makes reliable signal detection difficult for the base station, especially for the users at the cell edges. This issue has been actually recognized in 5G, and as a response, the 3rd generation partnership project (3GPP) has approved a work plan for enhancing the coverage of the physical random access channel for its Release 18 \cite{9927255}.

\begin{figure*}[h!]
\centering
\subfloat[Cellular Networks]
{\label{cellular}\includegraphics[width=0.33\textwidth]{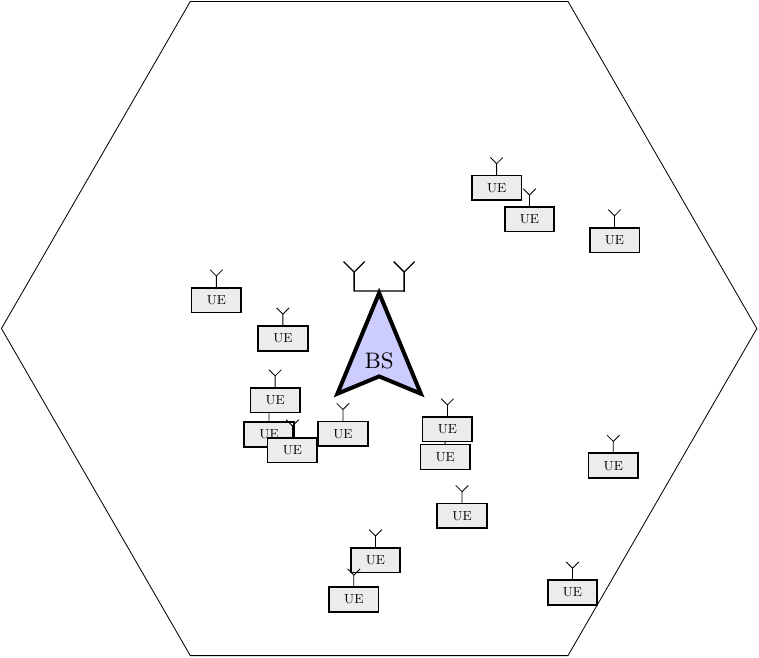}}
\hfill
\subfloat[Small-Cell Networks]
{\label{smallcell}\includegraphics[width=0.3\textwidth]{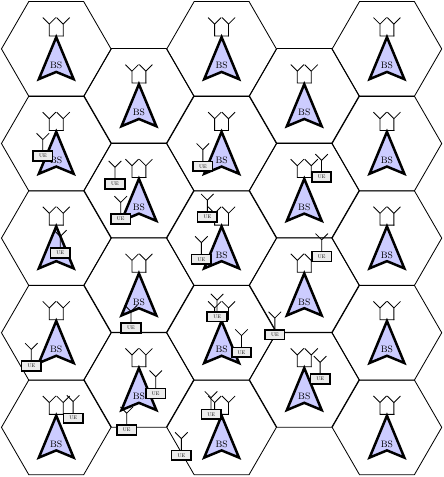}}
\hfill
\subfloat[Cell-Free Networks]
{\label{cf}\includegraphics[width=0.33\textwidth]{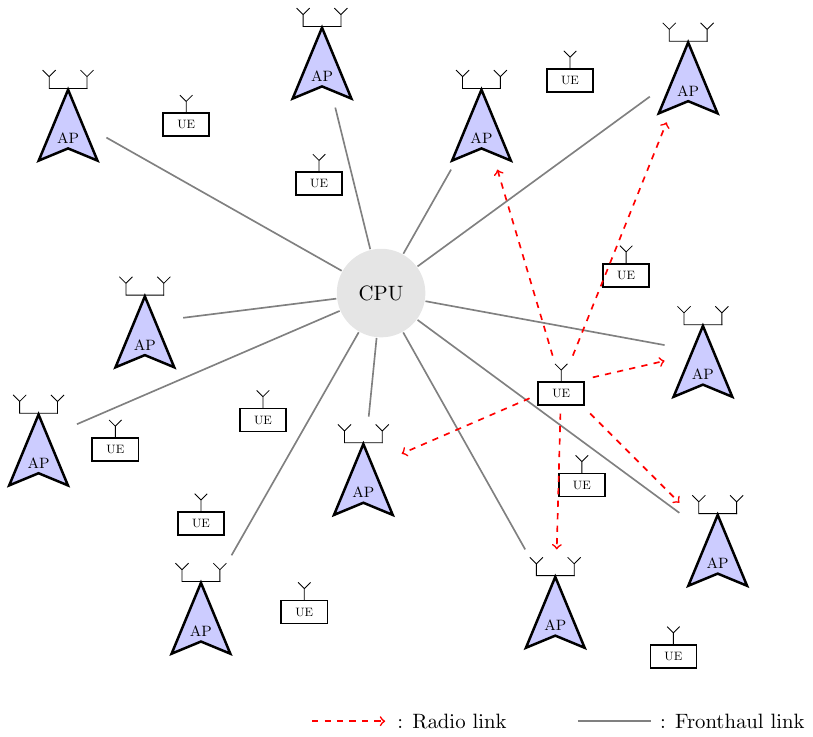}}
\caption{Comparison between cellular networks, small-cell networks and user-centric cell-free networks. In our study, we ensure a fair comparison by maintaining the same number of base station antennas in each network system and an equal mean of active users over the same area size.}
\label{FigSystemModel}
\end{figure*}

Meanwhile, the recent development in wireless communication has shown that the newly emerging networks, known as \textit{cell-free} massive MIMO, are able to effectively provide almost uniformly service over the coverage area \cite{7827017, 7917284}. Compared to cellular networks, there are no clear boundaries dictating the service area of a particular base station/access point (AP) as depicted in Fig. \ref{cf}. Instead in cell-free networks, a user can be served by a large number of distributed APs which are connected to a central processing unit (CPU) via fronthaul links. Due to the distributed deployment of APs, it is apparent that the distance between users and APs becomes closer. Thus, cell-free networks can potentially alleviate the coverage problem that mainly arises in low-power IoT applications.

Motivated by those problems, we investigate in this paper the \textit{user-centric} scheme for cell-free massive MIMO in supporting the IoT application. This approach is different from the conventional \textit{network-centric} approach, in which non-overlapping service areas are determined according to the network parameters. This can lead to poor performance for users at the edges. A user-centric approach instead allows users to be served by a subset of network APs that are chosen based on the user's characteristics. To achieve this, a collaboration of multiple APs is required to create user-specific overlapping service areas. 
The goal of this paper is to investigate how well the user-centric cell-free massive MIMO can perform compared to other networks when using a simple slotted ALOHA in grant-free random access protocol and to identify which parameters have the most impact on performance.

A considerable amount of work has been done on slotted ALOHA, specifically in the context of grant-free random access protocol. In \cite{5668922}, the authors proposed an irregular repetition slotted ALOHA (IRSA) scheme, which has significantly improved the efficiency of slotted ALOHA. By taking advantage of forward error correction and successive interference cancellation (SIC), this scheme has been extended in \cite{7302046}, referred to as coded slotted ALOHA (CS-ALOHA). A T-fold ALOHA scheme with lower complexity coding and lower required energy-per-bit for each user is proposed in \cite{8006985}. Assuming all users share a common codebook, a novel framework of \textit{unsourced} random access \cite{8006984} has emerged and has been further studied in \cite{8885472, 9153051, 9374476, 9432925}.

Nevertheless, to the best of the authors' knowledge, there has been only limited study investigating slotted ALOHA for user-centric cell-free massive MIMO. The most related works in the context of grant-free random access for cell-free massive MIMO are given in \cite{9712617, 9507457}. However, they investigated different schemes and did not make a clear-cut comparison to other networks. Thus in this paper, we attempt to fill the gap. Particularly, the main contribution of this paper is a comparative study between user-centric cell-free massive MIMO and other types of networks in terms of the throughput performance of slotted ALOHA. The choice of slotted ALOHA scheme in this paper is due to its simplicity, whereas the focus of this paper is more on the network side. Subsequently, in Section \ref{SystemModel}, we describe the system model including the channel and the slotted ALOHA model. Based on this model, we provide in Section \ref{SystemTput} the random access scheme that we use for user-centric cell-free massive MIMO and express its sum-throughput. It is then followed by numerical evaluation in Section \ref{numResults}, from which we can analyse the performance of various networks, and draw the conclusion of the paper in Section \ref{Conclusion}.

\section{SYSTEM MODEL}\label{SystemModel}

Considering the different types of networks shown in Fig. \ref{FigSystemModel}, this section focuses on the model of a user-centric cell-free massive MIMO system depicted in Fig. \ref{cf} in the uplink direction, where $K$ potential single-antenna users should be served by $L$ APs each of which is equipped with $N\geq 1$ antennas giving the total APs' antenna in the system $M=LN$. The users and the APs are distributed over a large area of size $A$. We consider a random access transmission, where only $K_a$ out of $K$ users are attempting to send their messages to the overlapping AP subsets $\mathcal{M}_k \subset \{1, \dots, L\}$ for $k=1, \dots, K_a$. The received messages at the APs are then forwarded to a CPU over error-free fronthaul links. In the case of standard cell-free massive MIMO, we have $\mathcal{M}_k = \{1, \dots, L\}$, where all APs should serve all the active users. In this paper, the set $\mathcal{M}_k$ for $k=1, \dots, K_a$ is assumed to be fixed on a large scale and known wherever it is needed.

Further, we use the dynamic cooperation clustering (DCC) framework given in \cite{SIG-109} in order to model which users are served by which APs. We accordingly introduce a set of diagonal matrices $\mathbf{D}_{kl} \in \mathbb{C}^{N\times N}$, for $k=1, \dots, K$ and $l=1, \dots, L$, where we define
\begin{align}
    \mathbf{D}_{kl} = 
    \begin{cases}
    \mathbf{I}_N & l\in \mathcal{M}_k \\
    \mathbf{0}_{N} & l\notin \mathcal{M}_k.
    \end{cases}
\end{align}
The matrix $\mathbf{D}_{kl}$ is used by AP $l$ to choose among the active users $K_a$ that the AP $l$ may serve. If the user $k$ has the index $l$ in its AP subset $\mathcal{M}_k$, then the matrix $\mathbf{D}_{kl}$ is an identity matrix, and the AP $l$ may consider user $k$ transmitting a data packet. Otherwise, the matrix $\mathbf{D}_{kl}$ is zeros, and any data transmission from user $k$ is neglected at AP $l$, disregarding there is no packet collision. In this case, each user is associated with a subset of APs that form a specific cluster. The user association or cluster formation can practically be done for example by a three-step procedure as proposed in \cite{9064545}. Suppose that 
\begin{align}
    \mathcal{D}_l=\{k : \mathrm{tr}(\mathbf{D}_{kl})\geq 1, k \in\{1, \dots, K \} \}
\end{align}
is the subset of UEs connected to at least one antenna at the AP $l$. The procedure begins with the user $k$ appointing a master AP by measuring the smallest path loss from the broadcasted synchronization signal. The appointed master AP responds by assigning a pilot and informing its neighbouring APs to serve the user $k$. Based on some rules, the neighbouring APs then decide if the user $k$ can be included in $\mathcal{D}_l$. 
In this paper, we assume that each user is a member of at least one subset $\mathcal{D}_l$ for $l=1, \dots, L$.

\subsection{Channel Model}
 
The channel follows the standard block fading model, where it is assumed that the channel vectors are time-frequency invariant during the so-called \emph{coherence block} and changing independently at random from block to block. We denote the channel in an arbitrary coherence block between UE $k$ and AP $l$ by a vector $\mathbf{g}_{kl}\in \mathbb{C}^N$ written as
\begin{equation}
    \mathbf{g}_{kl}=\sqrt{\beta_{kl}}\mathbf{h}_{kl},
\end{equation}
where $\mathbf{h}_{kl}\sim \mathcal{CN}(\mathbf{0}_N,\mathbf{I}_N)$ is the vector of small scale fading coefficients, and $\beta_{kkl}$ denotes the path loss.
The small-scale fading coefficients are modelled as uncorrelated Rayleigh fading, whereas the path loss is modelled as a function of propagation distance $d_{kl}$ between UE $k$ and AP $l$. In this paper, we choose 
\begin{equation}
    \beta_{kl}=-30.5-36.7 \,\textrm{log}_{10}\left(\frac{d_{kl}}{1 \textrm{m}}\right) \quad [\textrm{dB}].
\end{equation}
 We assume that the CPU knows the perfect channel state information (CSI) for the purpose of performing coherent data detection of the users. 
 To acquire the CSI, the data transmission is preceded by the transmission of orthogonal pilot sequences. The channel is then estimated from the received pilots at the CPU. To concentrate only on the performance of random data transmission, we assume that the CPU is able to estimate the channel exactly.

\subsection{The Slotted ALOHA System Model}
We consider that part of the signal dimension $\tau_c$ fitting with a coherence block is split into a length of $\tau_p$ and $\tau_d$, respectively for pilot signal and data transmission slot. The slotted transmission allows the CPU to perform a coherent reception from many APs. Accordingly, the users can send their data packet in the data transmission slot in a sporadic but coarsely synchronized fashion. In this case, the slotted ALOHA scheme is used in the access link between the users and the APs, whereas an orthogonal scheme is applied in the fronthaul links. Due to this assumption, the packet collision can only occur in the access link.
To make it straightforward, we assume that there is no delay caused by the transmission in the fronthaul links and by the processing at the APs.

Further, we assume that the packet that arrives at each user is independent of other users. Once a packet arrives at any user, the user becomes active at any random point in a finite space and sends directly its packet to the given slot. In this case, the packet traffic and the user activity are described by a single process. Thus, we model the number of active users in a given slot as a random variable $K_a \sim \textrm{Bin}(1, \pi)$, which has a binomial distribution with activation probability $\pi$. The mean value of active users in a certain slot is then given as $\mathbb{E}\{K_a\}=\pi K$. Whenever more than one users send data in the same slot, there is a collision. However, all of the colliding packets are not necessarily dropped. In this case, we consider a slotted ALOHA with capture effect \cite{DBLP:journals/winet/Zorzi98}. Depending on the channel condition, the packet with stronger signal power than other interfering packets has the possibility to be captured by the receiver. In the case that a user is not successful sending his packet, this user will be in a backlogged state.
We consider that the capture effect holds at the CPU, whereas the APs just forward the colliding packets to the CPU. The model used to describe the capture capability is based on the signal-to-interference-noise ratio (SINR). If the SINR of a particular user $k$ is greater than a required threshold $\alpha$, then the packet can be successfully decoded. Thus, we can express the packet success probability of user $k$ as
\begin{align}
    P_k(\alpha)=P\left[\textrm{SINR}_k > \alpha\right]
\end{align}

\section{SYSTEM THROUGHPUT} \label{SystemTput}

With the given model in the previous section, we provide in this section the throughput of slotted ALOHA for the user-centric cell-free massive MIMO. To this end, we look first at the SINR of user $k$ given only $K_a$ active users from the total $K$ potential users. In this setting, the received signal observed at a particular AP $l$ can be expressed as
\begin{equation}
    \mathbf{y}_l= \sum_{k=1}^{K_a} \mathbf{g}_{kl}s_k+\mathbf{n}_l,
\end{equation}
where $s_k\in \mathbb{C}$ is the data symbol transmitted by the active user $k$ with power $ p_k= \mathbb{E}\{\lvert s_k \rvert ^2\}$ and $\mathbf{n}_l \sim \mathcal{CN}(\mathbf{0}_N, \sigma_l^2\mathbf{I}_N)$ is an independent noise vector at the AP $l$.  

To detect the data symbol $s_k$, the AP $l$ can apply a received combining vector $\mathbf{v}_{kl}$ together with the assignment matrix $\mathbf{D}_{kl}$ to form an effective combining vector $\mathbf{D}_{kl}\mathbf{v}_{kl}$.
Then, the AP $l$ computes the inner product between the effective combining vector $\mathbf{D}_{kl}\mathbf{v}_{kl}$ and the received signal $\mathbf{y}_l$. As a result, we obtain the symbol estimate of user $k$ at the AP $l$ given by
\begin{align}
    \widehat{s}_{kl}=\mathbf{v}_{kl}^{\textrm{H}}\mathbf{D}_{kl}\mathbf{y}_l.
    \label{innerproduct_l}
\end{align}
Hence, only the APs from the set $\mathcal{M}_k$ are assigned to perform the detection.
Adopting the fully centralized operation \cite{7917284, 8845768, 8730536}, where the APs should only forward the received baseband signal $y_l$, the symbol estimate $\widehat{s}_{k}$ can then be computed at the CPU by summing up the inner product in (\ref{innerproduct_l}) for all considered APs expressed by
\begin{align}
    \widehat{s}_{k}&=\sum_{l=1}^{L}\mathbf{v}_{kl}^{\textrm{H}}\mathbf{D}_{kl}\mathbf{y}_l\\
    &=\mathbf{v}_{k}^{\textrm{H}}\mathbf{D}_{k}\mathbf{y},
    \label{S_hat_k}
\end{align}
where $\mathbf{v}_k=\left[ \mathbf{v}_{k1}^{\mathrm{T}}, \dots, \mathbf{v}_{kL}^{\mathrm{T}}\right]^{\mathrm{T}}\in \mathbb{C}^{LN}$ is the centralized combining vector, $\mathbf{D}_k=\mathrm{diag}\left(\mathbf{D}_{k1}, \dots, \mathbf{D}_{kl}\right)$ is a block-diagonal matrix, and $\mathbf{y}\in \mathbb{C}^{LN}$ is the collection of the received signal at the CPU written as
\begin{align}
    \mathbf{y}&=\begin{bmatrix}
           \mathbf{y}_{1} \\
           \vdots \\
           \mathbf{y}_{L}
         \end{bmatrix}
         = \sum_{k=1}^{K_a} \mathbf{g}_k s_k+\mathbf{n}.
         \label{aggregate_y_l}
\end{align}
In (\ref{aggregate_y_l}), we denote the collection of channel vector of user $k$ to all APs by $\mathbf{g}_k=\left[\mathbf{g}_{k1}^{\mathrm{T}}, \dots, \mathbf{g}_{kL}^{\mathrm{T}}\right]^{\mathrm{T}}$ and the collection of noise vector by $\mathbf{n}=\left[\mathbf{n}_{1}^{\mathrm{T}}, \dots, \mathbf{n}_{L}^{\mathrm{T}}\right]^{\mathrm{T}}$. By plugging (\ref{aggregate_y_l}) in (\ref{S_hat_k}), we obtain 
\begin{align}
    \widehat{s}_{k}=\mathbf{v}_{k}^{\textrm{H}}\mathbf{D}_{k}\mathbf{g}_k s_k+\sum\limits_{\substack{i=1 \\ i\neq k}}^{K_a} \mathbf{v}_{k}^{\textrm{H}}\mathbf{D}_{k}\mathbf{g}_i s_i+\mathbf{v}_{k}^{\textrm{H}}\mathbf{D}_{k}\mathbf{n},
\end{align}
where the first term constitutes the desired signal, the second term represents interference from other active users, and  the third term represents noise. Thus, the instantaneous SINR of user $k$ can be expressed as
\begin{align}
    \mathrm{SINR}_k=\frac{p_k \lvert \mathbf{v}_{k}^{\textrm{H}}\mathbf{D}_{k}\mathbf{g}_k \rvert^2}{\sum\limits_{\substack{i=1 \\ i\neq k}}^{K_a}p_i\lvert \mathbf{v}_{k}^{\textrm{H}}\mathbf{D}_{k}\mathbf{g}_i\rvert^2+\sigma^2\Vert \mathbf{D}_{k}\mathbf{v}_{k} \Vert^2}.
\end{align}
Due to its ability to maximize SINR, it is appealing to us to use the minimum mean-squared error (MMSE) combining vector given by 
\begin{align}
    \mathbf{v}_k^{\mathrm{MMSE}}=p_k\left(\sum_{i=1}^{K_a}p_i \mathbf{D}_{k}\mathbf{g}_i \mathbf{g}_i^{\mathrm{H}}\mathbf{D}_{k}+\sigma^2\mathbf{I}_{M} \right)^{-1}\mathbf{D}_{k}\mathbf{g}_k. 
    \label{MMSE_combining_vector}
\end{align}
It is to note that the realization of $K_a$ as well as their perfect CSI, their cluster formation and the power allocation are known at the CPU obtained from the preceding activity detection, CSI acquisition and cluster formation procedure, such that the computation of the MMSE combining vector in (\ref{MMSE_combining_vector}) is possible. Applying the combining vector (\ref{MMSE_combining_vector}), we obtain the SINR given by
\begin{align}    \mathrm{SINR}_k=p_k\mathbf{g}_k^{\mathrm{H}}\mathbf{D}_k\left(\sum\limits_{\substack{i=1 \\ i\neq k}}^{K_a}p_i \mathbf{D}_{k}\mathbf{g}_i \mathbf{g}_i^{\mathrm{H}}\mathbf{D}_{k}+\sigma^2\mathbf{I}_{M} \right)^{-1}\mathbf{D}_{k}\mathbf{g}_k.
        \label{SINR uc-cf}
\end{align}
In the case of standard cell-free massive MIMO, we have then an identity matrix for $\mathbf{D}_k$ in the SINR expression in (\ref{SINR uc-cf}).
Next, we are interested in the sum-throughput which is defined as the total number of packets from all users that are successfully decoded by the CPU given as
\begin{align}
    S_{\textrm{tot}}=\frac{\tau_d}{\tau_c}2B\,\sum_{k=1}^{K_a}\mathrm{log}_2\left(1+\mathbbm{1}_{(\alpha, \infty)}(\mathrm{SINR}_k)\, \mathrm{SINR}_k\right),
\end{align}
where $B$ is the available bandwidth and $\mathbbm{1}_{(\alpha, \infty)}(x)$ is an indicator function giving the value $1$ if $x\in {(\alpha, \infty)}$ and zero otherwise. Later in Section \ref{numResults}, we will numerically compare the sum-throughput of user-centric cell-free massive MIMO with cellular massive MIMO and small-cell networks, whose SINR can be given respectively by \cite{SIG-109}
\begin{equation}
\begin{aligned}
\operatorname{SINR}_{k}^{\mathrm{MIMO}} &= p_k \mathbf{g}_{k}^{\mathrm{H}}\left(p \sum_{\substack{i=1 \\
i \neq k}}^{K_a} \mathbf{g}_{i} \mathbf{g}_{i}^{\mathrm{H}} +\sigma^{2} \mathbf{I}_{M}\right)^{-1} \mathbf{g}_{k},
\end{aligned}
\label{SINR_cellularMIMO}
\end{equation}
and
\begin{equation}
\operatorname{SINR}_{k l}^{\text {small-cell }}=\frac{p_k\left|g_{k l}\right|^{2}}{p_k \sum_{\substack{i=1 \\ i \neq k}}^{K_a}\left|g_{i l}\right|^{2}+\sigma^{2}}.
\label{SINR_smallcell}
\end{equation}
We may observe from (\ref{SINR_cellularMIMO}) and (\ref{SINR_smallcell}) that these networks are the two extreme cases of cell-free massive MIMO in terms of how the processing is performed at the  receiver end. In the case of cellular massive MIMO, the signal from all antennas can be processed jointly, but the receiving antennas are co-located. On the other hand, the antennas in small cell are distributed, but the received signal of each antenna is processed separately.


\section{NUMERICAL RESULTS}\label{numResults}

In this section, we provide the performance evaluation of slotted ALOHA when it is applied in user-centric cell-free massive MIMO. An area of size $A=1 \mathrm{km}^2$ is considered in which the CPU is placed in the origin and surrounded by uniformly distributed APs. In the same area $A$, we also place $K=200$ users with uniform distribution, but only $K_a$ among them are active with probability $\pi$. In this scenario, the boundary effect is avoided by wrapping around the area at the edges. To evaluate the sum-throughput, we run a series of simulations assuming the available bandwidth $B=1$ MHz, the noise power $\sigma^2=-109$ dBm, the length of data $\tau_d=10$ symbols, and the coherence block of $\tau_c=20$ symbols. Further, we set at the receiver the threshold $\alpha=3$ dBm throughout the simulation. All these parameters should also apply to all standard networks we use later for benchmarking. Unless otherwise stated, the total number of APs' antennas should be the same for all networks given by $M=64$. In our simulation, it is assumed that the active users concurrently send their data packets in a single slot without retransmissions in every instance. Subsequently, the sum-throughput is obtained by averaging over many traffic and channel realizations.

\begin{figure}
    \centering
    \includegraphics[width=\columnwidth]{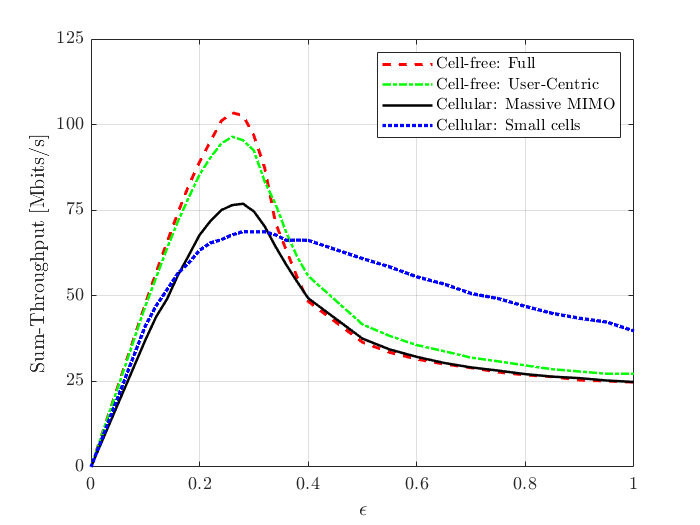}
    \caption{The sum-throughput of various networks as a function of activation probability $\pi$ for $L=16$ and $N=4$}
    \label{EpsilonM16N4_dcc}
\end{figure}

First, we investigate the sum-throughput of slotted ALOHA in relation to the activation probability $\pi$. We consider four types of networks for comparison, namely cellular massive MIMO networks, cellular small-cell networks, standard cell-free massive MIMO networks with the full set of APs, and user-centric cell-free massive MIMO with $\vert \mathcal{M}_k\vert=4$. We make use of the users' location to determine the subset $\mathcal{M}_k$ in the simulation, where the user $k$ is connected to 4 nearest located APs. The corresponding numerical results are shown in Fig. \ref{EpsilonM16N4_dcc} for the setup where the cell-free and small-cell networks consist of $L=64$ number of APs, each of which is equipped with single antenna $N=4$. In the case of cellular massive MIMO, the setup constitutes to a single AP (aka base station) with 64 co-located antennas.

As depicted in Fig. \ref{EpsilonM16N4_dcc}, the sum-throughput increases up to a certain value as the activation probability get higher. However, when the activation probability increases further, the sum-throughput gets decreases, and then tends to flatten as approaching $\pi=1$. This condition is as expected, since the higher the probability of active users the larger amount of packets are sent, which may lead to a collision that can no longer be captured by the receiver. Further observing Fig. \ref{EpsilonM16N4_dcc}, in fact, no single network can outperform the other networks over all ranges of $\pi$. In the regime of high activation probability $\pi> 0.4$, the small-cell networks are shown to have the highest sum-throughput.

\begin{figure} [t]
    \centering
    \includegraphics[width=\columnwidth]{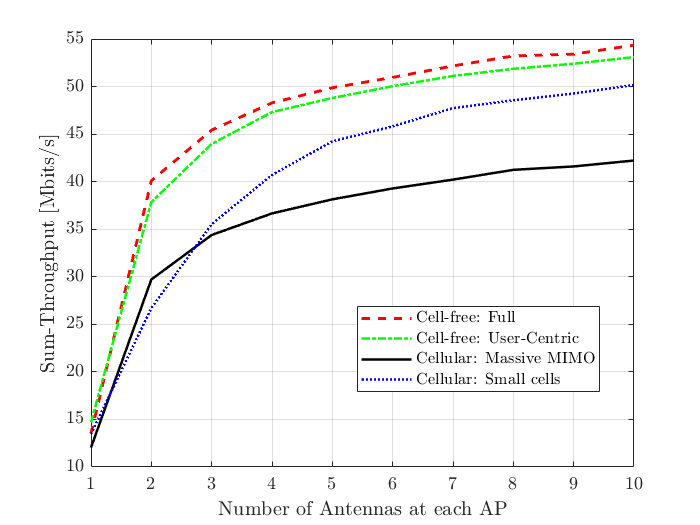}
    \caption{The sum-throughput with respect to the number of antennas at each AP  $N$ for $\pi=0.1$, $L=16$.}
    \label{M16Nepsilon1comma0}
\end{figure}

In contrast, the small-cell networks have the lowest sum-throughput in the regime of low $\pi$. The highest sum-throughput is achieved in this case by full cell-free networks and then followed by user-centric cell-free massive MIMO networks. Since the user traffic has a low $\pi$ in the typical IoT scenario, these results suggest us to deploy cell-free massive MIMO networks when IoT-type application is considered and a large sum-throughput is expected. Interestingly, for sporadic traffic, in which $\pi\ll 1$ is often the case, the performance gap between full and user-centric cell-free massive MIMO is almost negligible, while the user-centric networks require a lower fronthaul load. 

 To investigate further, we show in Fig. \ref{M16Nepsilon1comma0} the behaviour of the sum-throughput against the number of antennas at each AP for $\pi=0.1$ and $L=16$. In this setup, as we increase the number of antennas at each AP, the total number of antennas $M$ increases at the same time. To note here, the number of AP for cellular massive MIMO is always one. Thus, we multiply every increase in the number of antennas by $L$ for cellular massive MIMO in order to make a fair comparison to other networks. As shown in Fig. \ref{M16Nepsilon1comma0}, the performance of user-centric cell-free massive MIMO gets very close to the full cell-free massive MIMO all over the number of antennas. It can also be confirmed in Fig. \ref{M16Nepsilon1comma0}, that the small-cell networks outperform cellular massive MIMO in the case of large $N$.

\begin{figure}[t]
    \centering
    \includegraphics[width=\columnwidth]{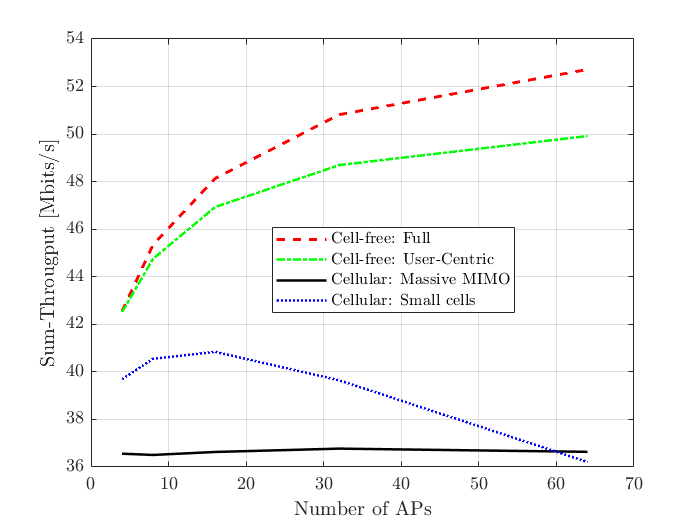}
    \caption{The sum-throughput with respect to the number of APs $L$ for $\pi=0.1$, $L=16$.}
    \label{MNepsilon1comma0_2000}
\end{figure}

Lastly, we investigate the sum-throughput in relation to the AP distribution at a low $\pi$ regime. For this purpose, we plot in Fig. \ref{MNepsilon1comma0_2000} the sum-throughput against the number of APs  while keeping the total number of antennas in the networks fixed. As can be observed, the sum-throughput increases proportionally to the spread of antennas for full and user-centric massive MIMO. Together with Fig. \ref{M16Nepsilon1comma0} this means, that higher performance can always be achieved for cell-free massive MIMO by increasing the number of antennas at APs and by allowing them to be more distributed in the given service area. However, a different trend is shown by small-cell networks, where the sum-throughput only increases at a small number of APs, but then decreases with more distributed antennas. This result can be explained by the fact that the distributed antennas in cell-free massive MIMO can cooperate to manage interference. In contrast, the small-cell networks do not exploit the feature of distributed antennas to capture a possible packet collision. Nevertheless, the distributed antennas in small-cell networks can still help in dividing the service location such that a collision can be avoided, especially for far-apart users. This might be the reason why they have a better performance than cellular massive MIMO networks.

\section{CONCLUSION}\label{Conclusion}
In this paper, the performance of slotted ALOHA in terms of its sum-throughput was studied for user-centric cell-free massive MIMO and compared with several types of networks. In particular, we have provided the sum-throughput expression for user-centric cell-free massive MIMO when the slotted ALOHA scheme is applied. Further, we have numerically evaluated the sum-throughput with respect to some practical parameters such as activation probability, number of antennas per AP, and number of distributed APs. The study gives insights into the potential of user-centric massive MIMO to be utilized for IoT applications which are heavily characterized by sporadic traffic, low power and the need for wide coverage. Our simulation results revealed, that the full cell-free massive MIMO is in general the obvious choice when it comes to the wide coverage, high throughput, and sporadic traffic with activation probability $\pi \ll 1$ as typically found in many IoT applications. On the other hand, user-centric cell-free massive MIMO offers an interesting alternative. It doesn't require all APs to participate in performing the detection, which results in a lower fronthaul load, but only has a negligible performance degradation, and still far better than small-cell networks and cellular massive MIMO. The results suggest the potential of user-centric cell-free massive MIMO to support the IoT application with a compromised feature between coverage, high throughput, and low fronthaul load.


%
%



\ifCLASSOPTIONcaptionsoff
  \newpage
\fi



\bibliographystyle{IEEEtran}
{\footnotesize \bibliography{References}}

%
%







\end{document}